%% file: main.tex
\documentclass[10pt,shortpaper,twoside,web]{ieeecolor}
\usepackage[utf8]{inputenc}
\usepackage[T1]{fontenc}
\usepackage{generic}
\usepackage{cite}
\usepackage{amsmath,amssymb,amsfonts}
\usepackage{algorithmic}
\usepackage{graphicx}
\usepackage{textcomp}
\usepackage{subfig}
\def\BibTeX{{\rm B\kern-.05em{\sc i\kern-.025em b}\kern-.08em
    T\kern-.1667em\lower.7ex\hbox{E}\kern-.125emX}}
\begin{document}
\title{Characterization and Analysis of On-Chip Microwave Passive Components at Cryogenic Temperatures}
\author{B.~Patra,~\IEEEmembership{Student~Member,~IEEE,}
        M.~Mehrpoo,~\IEEEmembership{Student~Member,~IEEE,}
        A.~Ruffino,~\IEEEmembership{Student~Member,~IEEE,}
        F.~Sebastiano,~\IEEEmembership{Senior~Member,~IEEE,}
        E.~Charbon,~\IEEEmembership{Fellow,~IEEE,}
        M.~Babaie,~\IEEEmembership{Member,~IEEE}
\thanks{This work was supported by Intel corporation.}
\thanks{B. Patra, M. Mehrpoo, F. Sebastiano and M. Babaie are with the Department of Quantum and Computer Engineering, Delft University of Technology, 2628 CD Delft, The Netherlands.}% <-this % stops a space
\thanks{A. Ruffino and E. Charbon are with the Institute of Microengineering, Faculty of Engineering, École Polytechnique Fédérale de Lausanne (EPFL), 2002 Neuchâtel, Switzerland.}}

\maketitle

\begin{abstract}
This paper presents the characterization of microwave passive components, including metal-oxide-metal (MoM) capacitors, transformers, and resonators,  at deep cryogenic temperature (4.2\,K). The variations in capacitance, inductance and  quality factor are explained in relation to the temperature dependence of the physical parameters and the resulting insights on modeling of passives at cryogenic temperatures are provided. Both characterization and modeling, reported for the first time down to 4.2\,K, are essential in designing cryogenic CMOS radio-frequency integrated circuits, a promising candidate to build the electronic interface for scalable quantum computers.
\end{abstract}


\input{introduction.tex}

\input{section2.tex}
\input{section3.tex}
\input{section4.tex}
\input{conclusion.tex}

\clearpage
\IEEEtriggeratref{13}
\bibliographystyle{IEEEtran}
\bibliography{main.bib}

\end{document}

%% file: introduction.tex
\section{Introduction}
\label{sec:section1}

Complementary Metal Oxide Semiconductor (CMOS) circuits operating at cryogenic temperatures (Cryo-CMOS) have been proposed to build the  scalable control electronics of quantum processors \cite{jssc_bpatra}\cite{jeroen}. Besides that, Cryo-CMOS technology has been used in the past to fabricate cryogenic low noise amplifiers (LNAs) for high sensitivity receivers \cite{cryo_LNA} and cryogenic LC oscillators for electron spin resonance detectors \cite{ESR}. The advent of cryo-CMOS has triggered the need for the characterization of active and passive components at cryogenic temperatures as required to reliably predict the performance of cryogenic radio frequency integrated circuits (RFIC). 
 
To some extent, this has been pursued by the scientific community in case of active devices, which is evident from papers that show DC characterization  \cite{jeds_rosario}\cite{cryo_28nm}, RF and noise characterization \cite{cryoRF_77K}, device mismatch \cite{mismatch_4K}\cite{mismatch_pascal}, small signal and noise characterization \cite{cryo_noise} of CMOS devices in different technology nodes. Although inductors have been characterized over the military temperature range \cite{temp_ind1}\cite{temp_ind2}, we have characterized, for the first time, a wide set of passive components at cryogenic temperatures.

%% file: section2.tex
\section{Test structures and measurement setup}
\label{sec:section2}

Several test structures were fabricated in the 1P7M-4x1Z1U TSMC 40-nm CMOS with an ultra-thick metal layer to characterize the passive components at 4\,K comprehensively.\\

A high-density metal-oxide-metal (MoM) capacitor was taped-out using stacked inter-digitated metal fingers in layers 1 to 5.
For the inductance characterization, a multi-turn transformer 
%featuring a two-turn primary with 190\,$\mu$m diameter and 8\,$\mu$m trace width and a two-turn secondary with 130~$\mu$m diameter and 7\,$\mu$m trace width 
was designed using the ultra-thick metal layer and avoiding substrate shielding, since high substrate resistivity is expected at 4\,K thanks to carrier freeze-out. 
%Shielding of the transformer was prevented due to a highly resistive substrate at 4\,K, thereby not having the need to reduce the tangential electric field losses in the low-resistive substrate \cite{ind_shield}.
Finally, a resonator was designed to validate the cryogenic model of the inductor and capacitor.
%and analyze its impact on cryogenic RFICs.

\begin{figure}[t]
\centering
\includegraphics[width=1\linewidth]{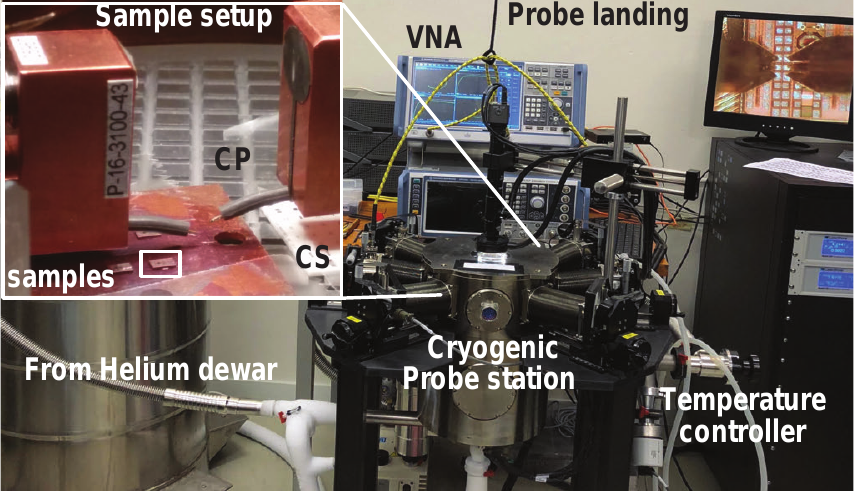}
\vspace{-.25in} 
\caption{Measurement setup}
\vspace{-.2in} 
\label{fig:1}
\end{figure}

% \begin{figure}[b]
%     \vspace{-.25in}
%     \subfloat[\label{fig:2a}]{
%     \includegraphics[width=0.3\textwidth]{figures/cap_model.eps}
%     }
%     \hfill
%     \subfloat[\label{fig:2(b)}]{
%     \includegraphics[width=0.15\textwidth]{figures/mom_micrograph.eps}
%     }
%     \caption{(a) MoM capacitor model and (b) micrograph.}
%     \label{fig:2}
% \end{figure}

\begin{figure}[b]
    \centering
    \vspace{-.2in}
    \includegraphics{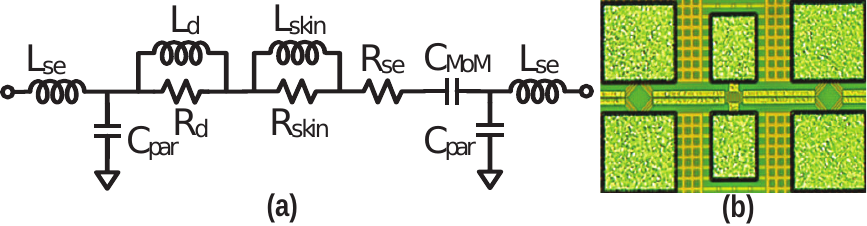}
    \caption{(a) MoM capacitor model and (b) micrograph.}
    \vspace{-.25in}
    \label{fig:2}
\end{figure}

\begin{figure*}
    \centering
    \vspace{-.3in}
    \includegraphics{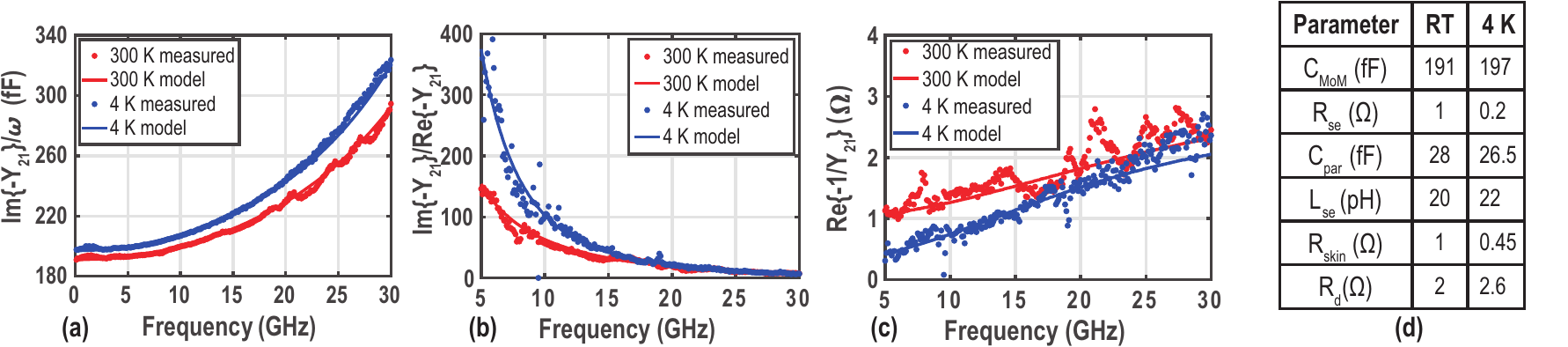}
    \caption{Extraction of: (a) Capacitance, (b) Quality factor, (c) Series resistance. (d) Model parameters at RT and 4\,K.}
    \vspace{-.25in}
    \label{fig:3}
\end{figure*}

% All the components were connected to ground-signal-ground (GSG) pads without electrostatic discharge (ESD) protection diodes and with low capacitance for probing with a 100 um GSG probe, as shown in the chip micrographs in Fig.\,\ref{fig:2}\,(b) , \ref{fig:4}\,(b), \ref{fig:5}\,(b). 
%The two ground pads in the GSG structure are shorted to each other at metal 1 level after using vias from AP to M1 layer, thus creating the signal to ground capacitance between M6 and M1.

% \begin{figure*}[t]
% \centering
% \includegraphics[width=1\linewidth]{figures/mom_v2.pdf}
% %\hspace{-0.25in}
% \vspace{-.25in} 
% \caption{(a) MoM model (b) Micrograph (c) -$Y_{12}/\omega$ (d) Quality factor (inset-series resistance)}
% \vspace{-.2in} 
% \label{fig:sec3_fig1}
% \end{figure*}

%Although the die temperature was not measured, considering the area under the device under test (DUT) to be large, no static power dissipation, large copper mass below the dies and sample holder containing the temperature sensor reading 4.2 K, it can be concluded that the DUT was at 4.2 K.

The measurements were done using ground-signal-ground (GSG) probes in a 40\,GHz Lakeshore CPX cryogenic probe station with R$\&$S ZNB40 vector network analyzer (VNA) (see Fig.\ref{fig:1}). To ensure proper thermalization, the dies were mounted  with a conductive glue on a copper plate (CP), which was securely taped to the sample holder. 
%The thermal conductivity of boron doped silicon layers at temperatures below 20\,K is almost the same as it is at 300\,K \cite{si_thermal_cond}, while thermal conductivity of copper remains the same or improve depending on its purity at 4\,K \cite{cte_copper}, suggesting good thermal link between sample holder and device under test. 
%Due to the variation in probe contact resistance caused by thermal fluctuation affecting the measurement at cryogenic temperatures, short-open-load-through (SOLT) calibrations of the probe tips were done right before the measurement using Picoprobe calibration substrate (CS) also mounted on the sample holder, at the measurement temperature.
Due to the variation in the GSG probe electrical characteristics at cryogenic temperatures, short-open-load-through (SOLT) calibrations were done right before the measurement using a Picoprobe calibration substrate (CS) also mounted on the sample holder, at the measurement temperature. 
%All the measurement results were de-embedded using open standard \cite{deembed} and then analyzed for performance variation in the following section.

%% file: section3.tex
\section{Measurement, Analysis and Modelling}
\label{sec:section3}
\subsection{MoM capacitor}
The MoM capacitor can be modeled as a $\pi$-network \cite{cap_model} as shown in Fig.\ref{fig:2}\,(a), where C$_{\textup{MoM}}$ is the actual capacitance due to the interdigitated metal fingers across an extra low-k inter-metal dielectric \cite{tsmc_dielectric}, and C$_{\textup{par}}$ represents the parasitic  between the lowest metal layer and the ground plane (poly shield to filter substrate noise). R$_{\textup{se}}$ and L$_{\textup{se}}$ represent the equivalent series resistance and inductance respectively, of the traces and vias from the  pad to the device terminals.
%caused by the lead/plate material losses, 
%with inter-level dielectrics (ILD) like silicon dioxide
The frequency dependent losses is modeled as R$_{\textup{skin}}$ and L$_{\textup{skin}}$ due to skin effect and  R$_{\textup{d}}$ and L$_{\textup{d}}$ due to the dielectric. Note that the quality factor of the capacitor above 10\,MHz is limited by the series resistance \cite{cap_temp} and hence, the leakage resistance (modeled as a very high resistance across the capacitor terminals at DC) due to the interface traps \cite{cap_leakage} is ignored in the model.
 
 Fig.\,\ref{fig:3}\,(a) shows the measured $Im\{-Y_{21}\}/\omega$ ($\omega$ is the angular frequency), from which the C$_{\textup{MoM}}$ can be extracted at the lowest measured frequency (i.e., 100\,MHz), where the effect of parasitic inductance is negligible \cite{cap_extraction}. Similarly, C$_{\textup{par}}$ can be extracted from measured $Y_{11}+Y_{21}$. Both the capacitances incur slight change at 4\,K compared to room temperature (RT) due to variation in the dielectric constant \cite{sio2_temp}, as the thermal contraction of metals is negligible \cite{cte_copper}.
 
%The quality factor of the capacitor above 10\,MHz is limited by the series resistance \cite{cap_temp} and hence, the leakage resistance (modeled as a very high resistance across the capacitor terminals at DC) due to the interface traps \cite{cap_leakage} is ignored in the model.
The measurement uncertainty increases when the desired impedance is negligible compared to the VNA reference impedance of 50\,$\Omega$. Hence, at the frequencies below 5\,GHz, the error in the determination of the series resistance and capacitor's quality factor would be significant and is excluded in Fig.\,\ref{fig:3}\,(b).
Due to reduction of dielectric and metal loss at lower temperatures, there is a boost in the quality factor at frequencies below 10\,GHz.
However, the dielectric loss does not improve over temperature above a certain frequency \cite{cap_temp}. Consequently, a negligible quality factor improvement is observed above 15\,GHz, when the dielectric loss is dominant, as can be gathered from Fig.\,\ref{fig:3}\,(b) and (c). Table I (Fig.\,\ref{fig:3}\,(d)) concludes the discussion on MoM capacitors and summarizes the change of the model parameters over temperature.

%$R_{MoM}$ models the leakage which is a relatively large-value resistor which arises due to non-ideal insulators used as dielectric, thus allowing some DC current to pass when a constant voltage is applied.

%The variation of capacitance over frequency is defined by the dissipation factor/loss tangent (imaginary part of permittivity) of the dielectric. The increase in capacitance suggests the increase in real permittivity and the increase in the rate of increase of capacitance over frequency suggests the increase in imaginary permittivity.

% \begin{equation}
% \begin{split}
%     Y11 &= C_{pad}+C_{MoM}+C_{ox}+C_{sub} \\
%     Y12 &= -C_{MoM} \\
%     Y11+Y12 &= C_{pad}+C_{ox}+C_{sub} 
% \end{split}
% \end{equation}
\vspace{-.1in}
\subsection{Transformer}

% \begin{figure*}[t]
% \centering
% \includegraphics[width=1\linewidth]{figures/xfmr_v2.pdf}
% \vspace{-.25in} 
% \caption{(a) Transformer model (b) Micrograph (c) Primary inductance (inset - coupling factor) (d) Quality factor (inset - series impedance)}
% \vspace{-.2in} 
% \label{fig:sec3_fig2}
% \end{figure*}

\begin{figure*}
    \centering
    \vspace{-.3in}
    \includegraphics{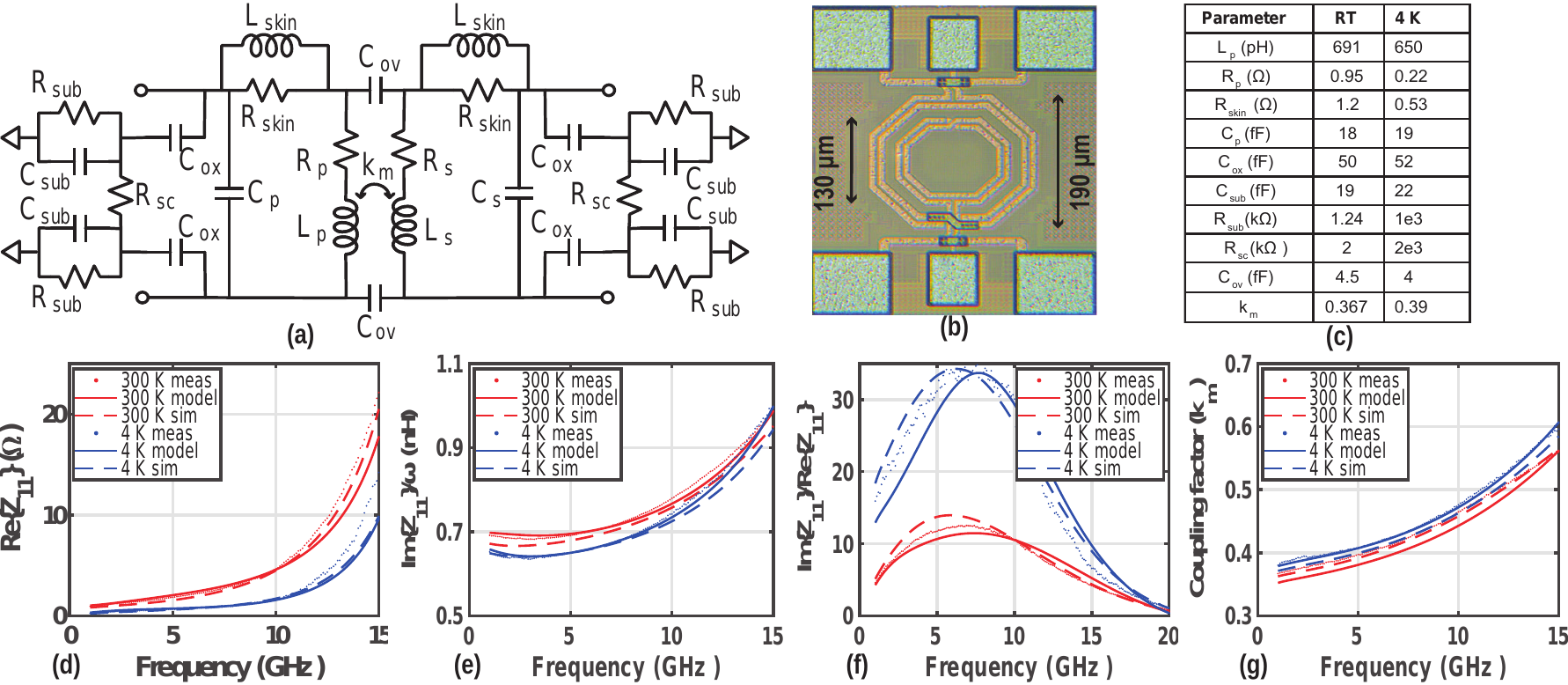}
    \caption{(a) Transformer model, (b) Micrograph and (c) Model parameters at RT and 4\,K. Extraction of: (d) Series resistance, (e) Inductance, (f) Quality factor and (g)  Coupling factor.}
    \vspace{-.25in}
    \label{fig:4}
\end{figure*}

The transformer can be modeled using the well-known frequency independent lumped model for on-chip spiral inductors \cite{ind_model_freq_ind} as depicted in Fig.\,\ref{fig:4}\,(a), where L$_\textup{p}$ and L$_\textup{s}$ represent the inductance, R$_\textup{p}$ and R$_\textup{s}$ describe the DC ohmic loss, of the primary and secondary windings, respectively. k$_\textup{m}$ represents the coupling factor of the transformer. C$_\textup{ov}$ models the interwinding capacitance, C$_{\textup{ox}}$ denotes the oxide capacitance, while C$_\textup{p}$ represents the capacitance due to metal lines running in parallel in the multi-turn primary winding. C$_{\textup{sub}}$ and R$_{\textup{sub}}$ model the substrate capacitance and resistance respectively.

%The measurement results of the transformer suggest that the winding's inductance (extracted from $Im\{Z_{11}\}$ at the lowest measured frequency) reduces by $\sim$5$\%$ and the quality factor (obtained as $Im\{Z_{11}\}/Re\{Z_{11}\}$) increases substantially at 4\,K, as shown in Fig.\,\ref{fig:4}\,(d) and (e), respectively. 
%To analyze this, we use the well-known frequency independent equivalent lumped model \cite{ind_model_freq_ind} for on-chip spiral inductors, as depicted in Fig.\,\ref{fig:4}\,(a).
R$_\textup{p}$ (extracted from $Re\{Z_{11}\}$ shown in Fig.\,\ref{fig:4}\,(d) at 1\,GHz where the skin effect is negligible) is $\sim$5$\times$ lower at 4\,K compared to RT, due to the increase in copper conductivity ($\sigma_{cu}$) \cite{cte_copper}. 
 %The DC resistance (R$\rm_p$) of the metal (extracted from $Re\{Z_{11}\}$ shown in the inset of Fig.\,\ref{fig:sec3_fig2}\,(d) at 100\,MHz\footnote{the skin depth is higher than the thickness of the metal at this frequency.}) is $\sim$5x lower at 4\,K compared to RT, due to increase in copper conductivity ($\sigma_{cu}$) \cite{thermal_cond}.% 
 %At higher frequencies, the skin depth reduces and eventually becomes smaller than the thickness of the metal. Hence, the skin effect becomes the dominant source of the inductor loss. The skin depth ($\delta$) can be approximated by $\sqrt{\frac{\rho}{\pi \mu f}}$, where $\rho$ is the resistivity in $\Omega$-m, $\mu$ is the permeability in H/m and $f$ is the frequency in Hz. Since the conductivity increases by 5$\times$, skin depth and thus the ac resistance decrease by $\sim\sqrt{5}$ \cite{ind_model}.
 At higher frequencies, the skin effect dominates and the loss becomes proportional to $1/\sigma_{cu}\delta$, in which  the skin depth $\delta=\sqrt{2}/\sqrt{\omega \mu \sigma_{cu}}$, where $\mu$ and $\sigma_{cu}$ represent the magnetic permeability and conductivity of the conductor. Since the conductivity increases by 5$\times$, skin depth and thus the inductor loss at higher frequencies decreases by $\sim$$\sqrt{5}$ \cite{ind_model}, as confirmed by Fig.\,\ref{fig:4}\,(d) for frequencies above 10\,GHz.
 %This results in a square-root improvement of losses as a function of  $\sigma_{cu}$ \cite{ind_model}, leading to only $\sim\sqrt{5}$ reduction in R$\rm_{s/p}$ at high frequencies.
 Furthermore, the increase in conductivity leads to a reduction in inductance by $\sim$5$\%$ \cite{temp_ind3}, also confirmed by the measured winding's inductance (extracted from $Im\{Z_{11}\}$ at the lowest measured frequency) as shown in Fig.\,\ref{fig:4}\,(e).
%The measured quality factor of the windings were slightly lower than the simulation results at RT. This is attributed to the exclusion of metal fill in EM simulations, which is required to satisfy the density rules and can slightly degrade the Q-factor of an inductor \cite{ind_fill1}. 

Fig.\,\ref{fig:4}\,(f) reveals that the inductor peak quality factor (extracted from $Im\{Z_{11}\}/Re\{Z_{11}\}$) increases by 2.7$\times$ from RT to 4\,K. The improvement is partially contributed (1.6$\times$ as verified from EM simulation) by the increase in conductivity and partly due to the reduction of tangential electric field losses in the silicon substrate, as it becomes highly resistive due to freeze-out.

Figure\,\ref{fig:4}\,(g) shows the measured k$_\textup{m}$, calculated as $k_\textup{m}=Im\{Z_{21}\}/\sqrt{Im\{Z_{11}\}\cdot Im\{Z_{22}\}}$, at both RT and 4\,K. The coupling factor is mainly set by the physical dimensions of the transformer, which barely change over temperature (i.e., $< 1\%$ as shown in \cite{cte_copper}). The slight increase in coupling factor at 4\,K is due to the change in capacitive coupling caused by the change in dielectric constant as mentioned earlier.

Table II (Fig.\,\ref{fig:4}\,(c)) summarizes the values of model parameters at RT and 4\,K. All the capacitors slightly change over temperature, which is also in line with the extracted MoM model. However, (R$_{\textup{sub}}$) and substrate coupling resistance (R$_{\textup{sc}}$) increases by 3 orders of magnitude at 4\,K mainly due to substrate freeze out \cite{substrate_model}. For low resistive substrates, the capacitance from the windings to the ground plane is dominated by C$_{\textup{ox}}$ \cite{substrate_res1, substrate_res2}, while for highly resistive substrates, the effective capacitance is lowered by C$_\textup{sub}$ in series with C$_\textup{ox}$, resulting in a slight increase in the frequency where peak quality factor occurs. The self-resonance frequency of the transformer increases by 5$\%$, due to the decrease in both inductance and overlap capacitance. 
Furthermore, as can be gathered from Fig.\,\ref{fig:4}\,(d)-(g), EM simulations confirm all abovementioned behaviors and accurately predict the performance at 4\,K; by increasing the copper conductivity by 5$\times$ and increasing the substrate resistivity by 1000, in the foundry metal stack.

%$R_{imag}$ which represents the electric coupling between lines through the conductive substrate, is approximately proportional to the resistance of the substrate below the inductor \cite{ind_model_freq_ind}.

% The resistance at 100 MHz of M7 inductor at RT is 0.445 and at 4 K is 0.090 there is a decrease in resitivity by 5x. 

% After increasing the conductivity of M7 by 5x, the inductance in simulation becomes is 448 pH. (reduction of 3.2%).

%% file: section4.tex
\section{Impact on RFICs}
\label{sec:section4}

% \begin{figure}[t]
% \centering
% \includegraphics[width=1\linewidth]{figures/tank.pdf}
% % \vspace{-.25in} 
% \caption{(a) Tank impedance, (b) Micrograph}
% \vspace{-.2in} 
% \label{fig:sec3_fig3}
% \end{figure}

\begin{figure}[b]
    \centering
    \vspace{-.0in}
    \includegraphics{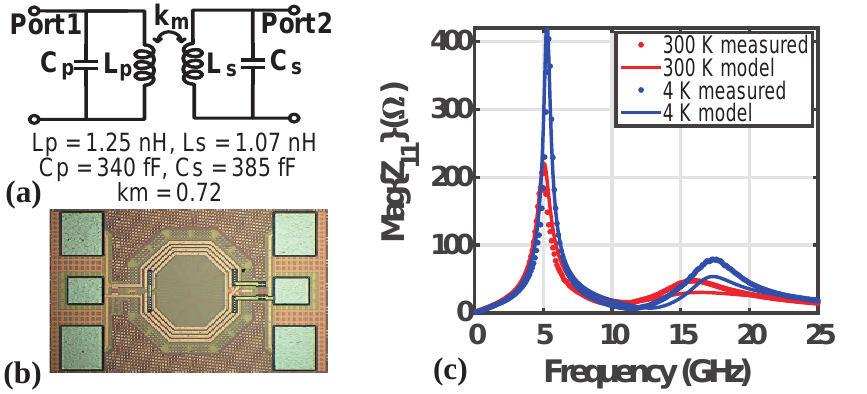}
    \caption{(a) Tank schematic, (b) micrograph and (c) input impedance.}
    \vspace{-.0in}
    \label{fig:5}
\end{figure}

%Transformers have been used in RFICs such as oscillators for low phase noise performance \cite{masoud}, receivers/transmitters for wideband input/output matching \cite{wideband_xfmr}, etc. 
To analyze the impact of the cryogenic operation on RFICs like oscillators \cite{masoud} and LNAs/PAs \cite{wideband_xfmr}, a transformer-based resonator (matching network), shown in Fig.\,\ref{fig:5}\,(a) was designed. Its performance was estimated using the developed models and compared with measurement results.
%The extracted resonator parameters are $L_p$ = 1.25\,nH, $L_s$ = 1.07\,nH, $k_m$ = 0.72, $C_p$ = 340\,fF, and $C_s$ = 385\,fF.
Fig.\,\ref{fig:5}\,(c) shows the input impedance of the resonator with Port2 open ($|Z_{11}|$), which determines an oscillator's power consumption in a transformer based oscillator\cite{masoud}.
There is an increase in the impedance peak of the resonator, from RT to 4\,K, due to the overall increase in quality factor, which is well predicted by the cryogenic model. Hence, one can obtain the same output voltage swing for half the current consumption, thus improving the oscillator's power efficiency.
The reduction in inductance and the effective parasitic capacitance causes the resonance to shift towards higher frequencies by 8\,$\%$. 
%Since $L_s.C_s/L_p.C_p \approx 1$, the ratio of the peak frequencies can be estimated as  $\sqrt{(1+k_m)/(1-k_m)}$ \cite{masoud}.
The ratio of the resonant frequencies merely depends on the coupling factor, which increases by 4$\%$ as predicted by the model.

%The frequency separation between the peaks in $|Z_{21}|$ at 4\,K can be estimated using transformer equations \cite{masoud}, which corresponds to the increase in $k_m$ at 4\,K, suggesting slightly wider band operation. 

%% file: conclusion.tex
\section{Conclusion}
\label{sec:conclusion}

Passive components at cryogenic temperatures  show  in general higher quality factor ($\sim$2$\times$) due to higher metal conductivity and lower loss in the substrate.
However, the value of inductive and capacitive on-chip components slightly changes (i.e., 5$\%$) from RT to 4\,K. Those variations can be replicated in an EM simulation by manipulating the resistivity of metals and substrate.
%The suggested variation in model parameters based on characterization, can be implemented by changing resistivity of metals and substrate in the metal stack provided by the foundry.
As a result, RFIC designers can predict the performance of cryogenic passive devices both by using EM simulation  and/or by scaling the presented lumped  model parameters. This enables, in combination with existing Cryo-CMOS models, the reliable design of cryogenic RFIC needed for future large-scale quantum computers.

%% file: main.bbl
\begin{thebibliography}{10}
\providecommand{\url}[1]{#1}
\csname url@samestyle\endcsname
\providecommand{\newblock}{\relax}
\providecommand{\bibinfo}[2]{#2}
\providecommand{\BIBentrySTDinterwordspacing}{\spaceskip=0pt\relax}
\providecommand{\BIBentryALTinterwordstretchfactor}{4}
\providecommand{\BIBentryALTinterwordspacing}{\spaceskip=\fontdimen2\font plus
\BIBentryALTinterwordstretchfactor\fontdimen3\font minus
  \fontdimen4\font\relax}
\providecommand{\BIBforeignlanguage}[2]{{%
\expandafter\ifx\csname l@#1\endcsname\relax
\typeout{** WARNING: IEEEtran.bst: No hyphenation pattern has been}%
\typeout{** loaded for the language `#1'. Using the pattern for}%
\typeout{** the default language instead.}%
\else
\language=\csname l@#1\endcsname
\fi
#2}}
\providecommand{\BIBdecl}{\relax}
\BIBdecl

\bibitem{jssc_bpatra}
B.~Patra, R.~M. Incandela, J.~P.~G. van Dijk, H.~A.~R. Homulle, L.~Song,
  M.~Shahmohammadi, R.~B. Staszewski, A.~Vladimirescu, M.~Babaie,
  F.~Sebastiano, and E.~Charbon, ``{Cryo-CMOS} circuits and systems for quantum
  computing applications,'' \emph{IEEE Journal of Solid-State Circuits},
  vol.~53, no.~1, pp. 309--321, Jan 2018, doi: \url{10.1109/JSSC.2017.2737549}.

\bibitem{jeroen}
J.~P. van Dijk, E.~Kawakami, R.~N. Schouten, M.~Veldhorst, L.~M. Vandersypen,
  M.~Babaie, E.~Charbon, and F.~Sebastiano, ``The impact of classical control
  electronics on qubit fidelity,'' \emph{arXiv preprint arXiv:1803.06176}, Mar
  2018.

\bibitem{cryo_LNA}
J.~Schleeh, G.~Alestig, J.~Halonen, A.~Malmros, B.~Nilsson, P.~A. Nilsson,
  J.~P. Starski, N.~Wadefalk, H.~Zirath, and J.~Grahn, ``Ultralow-power
  cryogenic {InP HEMT} with minimum noise temperature of {1 K} at {6 GHz},''
  \emph{IEEE Electron Device Letters}, vol.~33, no.~5, pp. 664--666, May 2012,
  doi: \url{10.1109/LED.2012.2187422}.

\bibitem{ESR}
G.~Gualco, J.~Anders, A.~Sienkiewicz, S.~Alberti, L.~Forró, and G.~Boero,
  ``Cryogenic single-chip electron spin resonance detector,'' \emph{Journal of
  Magnetic Resonance}, vol. 247, pp. 96 -- 103, October 2014, doi:
  \url"10.1016/j.jmr.2014.08.013".

\bibitem{jeds_rosario}
R.~M. Incandela, L.~Song, H.~Homulle, E.~Charbon, A.~Vladimirescu, and
  F.~Sebastiano, ``Characterization and compact modeling of nanometer {CMOS}
  transistors at deep-cryogenic temperatures,'' \emph{IEEE Journal of the
  Electron Devices Society}, vol.~6, pp. 996--1006, April 2018, doi:
  \url{10.1109/JEDS.2018.2821763}.

\bibitem{cryo_28nm}
A.~Beckers, F.~Jazaeri, A.~Ruffino, C.~Bruschini, A.~Baschirotto, and C.~Enz,
  ``Cryogenic characterization of 28 nm bulk {CMOS} technology for quantum
  computing,'' in \emph{2017 47th European Solid-State Device Research
  Conference (ESSDERC)}, Sept 2017, pp. 62--65, doi:
  \url{10.1109/ESSDERC.2017.8066592}.

\bibitem{cryoRF_77K}
A.~Siligaris, G.~Pailloncy, S.~Delcourt, R.~Valentin, S.~Lepilliet,
  F.~Danneville, D.~Gloria, and G.~Dambrine, ``High-frequency and noise
  performances of 65-nm {MOSFET} at liquid nitrogen temperature,'' \emph{IEEE
  Transactions on Electron Devices}, vol.~53, no.~8, pp. 1902--1908, Aug 2006,
  doi: \url{10.1109/TED.2006.877872}.

\bibitem{mismatch_4K}
N.~C. Dao, A.~E. Kass, C.~T. Jin, and P.~H.~W. Leong, ``Impact of series
  resistance on bulk {CMOS} current matching over the {5-300 K} temperature
  range,'' \emph{IEEE Electron Device Letters}, vol.~38, no.~7, pp. 847--850,
  July 2017, doi: \url{10.1109/LED.2017.2709545}.

\bibitem{mismatch_pascal}
P.~A. {'t Hart}, J.~P.~G. {van Dijk}, M.~{Babaie}, E.~{Charbon},
  A.~{Vladimircscu}, and F.~{Sebastiano}, ``Characterization and model
  validation of mismatch in nanometer {CMOS} at cryogenic temperatures,'' in
  \emph{2018 48th European Solid-State Device Research Conference (ESSDERC)},
  Sep. 2018, pp. 246--249, doi: \url{10.1109/ESSDERC.2018.8486859}.

\bibitem{cryo_noise}
A.~H. Coskun and J.~C. Bardin, ``Cryogenic small-signal and noise performance
  of 32nm {SOI CMOS},'' in \emph{2014 IEEE MTT-S International Microwave
  Symposium (IMS2014)}, June 2014, pp. 1--4, doi:
  \url{10.1109/MWSYM.2014.6848614}.

\bibitem{temp_ind1}
R.~Groves, D.~L. Harame, and D.~Jadus, ``Temperature dependence of {Q} and
  inductance in spiral inductors fabricated in a silicon-germanium/{BiCMOS}
  technology,'' \emph{IEEE Journal of Solid-State Circuits}, vol.~32, no.~9,
  pp. 1455--1459, Sept 1997, doi: \url{10.1109/4.628763}.

\bibitem{temp_ind2}
H.~. Chiu, Y.~. Lin, K.~Liu, and S.~. Lu, ``Temperature and substrate effects
  in monolithic {RF} inductors on silicon with 6-$\mu$m-thick top metal for
  {RFIC} applications,'' \emph{IEEE Transactions on Semiconductor
  Manufacturing}, vol.~19, no.~3, pp. 316--330, Aug 2006, doi:
  \url{10.1109/TSM.2006.879416}.

\bibitem{cap_model}
K.~{Lee}, S.~{Mohammadi}, P.~K. {Bhattacharya}, and L.~P.~B. {Katehi},
  ``Compact models based on transmission-line concept for integrated capacitors
  and inductors,'' \emph{IEEE Transactions on Microwave Theory and Techniques},
  vol.~54, no.~12, pp. 4141--4148, Dec 2006, doi:
  \url{10.1109/TMTT.2006.886157}.

\bibitem{tsmc_dielectric}
F.-W. Tsai, K.-C. Wang, K.-C. Lin, C.-L. Lin, and S.-M. Jeng, ``{Extreme}
  low-{K} dielectric film scheme for advanced interconnects,'' U.S. Patent
  7\,626\,245, Dec., 2009.

\bibitem{cap_temp}
P.~{Riess} and P.~{Baumgartner}, ``Temperature dependent dielectric absorption
  of {MIM} capacitors: {RF} characterization and modeling,'' in \emph{2006
  European Solid-State Device Research Conference}, Sep. 2006, pp. 459--462,
  doi: \url{10.1109/ESSDER.2006.307737}.

\bibitem{cap_leakage}
V.~F. {Tseng} and H.~{Xie}, ``Increased multilayer fabrication and {RF}
  characterization of a high-density stacked {MIM} capacitor based on selective
  etching,'' \emph{IEEE Transactions on Electron Devices}, vol.~61, no.~7, pp.
  2302--2308, July 2014, doi: \url{10.1109/TED.2014.2325491}.

\bibitem{cap_extraction}
H.~{Samavati}, A.~{Hajimiri}, A.~R. {Shahani}, G.~N. {Nasserbakht}, and T.~H.
  {Lee}, ``Fractal capacitors,'' \emph{IEEE Journal of Solid-State Circuits},
  vol.~33, no.~12, pp. 2035--2041, Dec 1998, doi: \url{10.1109/4.735545}.

\bibitem{sio2_temp}
F.~A. Miranda, W.~L. Gordon, V.~O. Heinen, B.~T. Ebihara, and K.~B. Bhasin,
  \emph{Measurements of Complex Permittivity of Microwave Substrates in the 20
  to 300 K Temperature Range from 26.5 to 40.0 GHz}.\hskip 1em plus 0.5em minus
  0.4em\relax Boston, MA: Springer US, 1990, pp. 1593--1599, doi:
  \url"10.1007/978-1-4613-0639-9-188".

\bibitem{cte_copper}
P.~Duthil, ``{Material Properties at Low Temperature},'' pp. 77--95. 18 p, Jan
  2015, doi: \url{10.5170/CERN-2014-005.77}.

\bibitem{ind_model_freq_ind}
Y.~Cao, R.~A. Groves, X.~Huang, N.~D. Zamdmer, J.~. Plouchart, R.~A. Wachnik,
  T.-J. King, and C.~Hu, ``Frequency-independent equivalent-circuit model for
  on-chip spiral inductors,'' \emph{IEEE Journal of Solid-State Circuits},
  vol.~38, no.~3, pp. 419--426, March 2003, doi:
  \url{10.1109/JSSC.2002.808285}.

\bibitem{ind_model}
C.~P. Yue and S.~S. Wong, ``Physical modeling of spiral inductors on silicon,''
  \emph{IEEE Transactions on Electron Devices}, vol.~47, no.~3, pp. 560--568,
  March 2000, doi: \url{10.1109/16.824729}.

\bibitem{temp_ind3}
S.~S. {Gerber}, ``Performance of high-frequency high-flux magnetic cores at
  cryogenic temperatures,'' in \emph{IECEC '02. 2002 37th Intersociety Energy
  Conversion Engineering Conference, 2002.}, July 2002, pp. 249--254, doi:
  \url{10.1109/IECEC.2002.1392019}.

\bibitem{substrate_model}
X.~{Huo}, P.~C.~H. {Chan}, K.~J. {Chen}, and H.~C. {Luong}, ``A physical model
  for on-chip spiral inductors with accurate substrate modeling,'' \emph{IEEE
  Transactions on Electron Devices}, vol.~53, no.~12, pp. 2942--2949, Dec 2006,
  doi: \url{10.1109/TED.2006.885091}.

\bibitem{substrate_res1}
D.~Eggert, P.~Huebler, A.~Huerrich, H.~Kueck, W.~Budde, and M.~Vorwerk, ``A
  {SOI-RF-CMOS} technology on high resistivity {SIMOX} substrates for microwave
  applications to 5 {GHz},'' \emph{IEEE Transactions on Electron Devices},
  vol.~44, no.~11, pp. 1981--1989, Nov 1997, doi: \url{10.1109/16.641369}.

\bibitem{substrate_res2}
R.~A. Johnson, C.~E. Chang, P.~M. Asbeck, M.~E. Wood, G.~A. Garcia, and
  I.~Lagnado, ``Comparison of microwave inductors fabricated on
  silicon-on-sapphire and bulk silicon,'' \emph{IEEE Microwave and Guided Wave
  Letters}, vol.~6, no.~9, pp. 323--325, Sep. 1996, doi:
  \url{10.1109/75.535833}.

\bibitem{masoud}
M.~{Babaie} and R.~B. {Staszewski}, ``A {Class-F CMOS} oscillator,'' \emph{IEEE
  Journal of Solid-State Circuits}, vol.~48, no.~12, pp. 3120--3133, Dec 2013,
  doi: \url{10.1109/JSSC.2013.2273823}.

\bibitem{wideband_xfmr}
M.~{Vigilante} and P.~{Reynaert}, ``On the design of wideband transformer-based
  fourth order matching networks for ${E}$ -band receivers in 28-nm cmos,''
  \emph{IEEE Journal of Solid-State Circuits}, vol.~52, no.~8, pp. 2071--2082,
  Aug 2017, doi: \url{10.1109/JSSC.2017.2690864}.

\end{thebibliography}
